\documentclass[modern]{aastex62}

\hypersetup{linkcolor=black,citecolor=black,filecolor=cyan,urlcolor=magenta}

\usepackage{graphicx}

\shorttitle{SOFIA Far Infrared Polarimetry}
\shortauthors{Jones et al.}

\begin{document}

\title{SOFIA Far Infrared Imaging Polarimetry of M82 and NGC 253: Exploring the Super--Galactic Wind}

\correspondingauthor{Terry Jay Jones}
\email{tjj@astro.umn.edu}

\author{Terry Jay Jones}
\affil{Minnesota Institute for Astrophysics \\
Universtity of Minnesota \\
116 Church St. SE, Minneapolis, MN 55455, USA}

\author{C. Darren Dowell}
\affil{NASA Jet Propulsion Laboratory, California Institute of Technology, 4800 Oak Grove Drive, Pasadena, CA 91109, USA}

\author{Enrique Lopez Rodriguez}
\affil{SOFIA Science Center/Universities Space Research Association}
\affil{NASA Ames Research Center, M.S. N232-12, Moffett Field, CA, 94035, USA}

\author{Ellen G. Zweibel}
\affil{Department of Astronomy, University of Wisconsin, Madison, WI 53706, USA}

\author{Marc Berthoud}
\affil{University of Chicago, Chicago, IL 60637, USA}
\affil{Yerkes Observatory, Williams Bay, WI, USA}

\author[0000-0003-0016-0533]{David T. Chuss}
\affil{Department of Physics, Villanova University, 800 E. Lancaster Ave., Villanova, PA 19085, USA}



\author[0000-0002-6622-8396]{Paul F. Goldsmith}
\affil{NASA Jet Propulsion Laboratory, California Institute of Technology, 4800 Oak Grove Drive, Pasadena, CA 91109, USA}

\author[0000-0001-6350-2209]{Ryan T. Hamilton}
\affil{Lowell Observatory, 1400 W Mars Hill Rd, Flagstaff, AZ 86001, USA}

\author[0000-0002-8702-6291]{Shaul Hanany}
\affil{School of Physics and Astronomy, University of Minnesota / Twin Cities, Minneapolis, MN, 55455, USA}

\author{Doyal A. Harper}
\affil{Department of Astronomy and Astrophysics, University of Chicago, Chicago, IL 60637, USA}
\affil{Yerkes Observatory, Williams Bay, WI, USA}

\author{Alex Lazarian}
\affil{Department of Astronomy, University of Wisconsin, Madison, WI 53706, USA}

\author[0000-0002-4540-6587]{Leslie W. Looney}
\affil{Department of Astronomy, University of Illinois, 1002 West Green Street, Urbana, IL 61801, USA}

\author[0000-0003-3503-3446]{Joseph M. Michail}
\affil{Department of Astrophysics and Planetary Science, Villanova University, 800 E. Lancaster Ave., Villanova, PA 19085, USA}
\affil{Department of Physics, Villanova University, 800 E. Lancaster Ave., Villanova, PA 19085, USA}

\author[0000-0002-6753-2066]{Mark R. Morris}
\affil{Department of Physics and Astronomy, University of California, Los Angeles, Box 951547, Los Angeles, CA 90095-1547 USA}

\author{Giles Novak}
\affil{Center for Interdisciplinary Exploration and Research in Astrophysics (CIERA), and Department of Physics \& Astronomy, Northwestern University, 2145 Sheridan Rd, Evanston, IL, 60208, USA}

\author[0000-0002-9650-3619]{Fabio P. Santos}
\affil{Max-Planck-Institute for Astronomy, K\"{o}nigstuhl 17, 69117 Heidelberg, Germany}

\author{Kartik Sheth}
\affil{NASA Headquarters, 300 E Street SW, DC 20546, USA}


\author{Gordon J. Stacey}
\affil{Department of Astronomy, Cornell University, Ithaca, NY 14853, USA}

\author{Johannes Staguhn}
\affil{Johns Hopkins University, Baltimore, Dpt. of Physics \& Astronomy, MD  21218, USA}
\affil{NASA Goddard Space Flight Center, Greenbelt, MD 20771, USA}

\author{Ian W. Stephens}
\affil{Harvard-Smithsonian Center for Astrophysics, 60 Garden Street, Cambridge, MA, USA}

\author{Konstantinos Tassis}
\affil{Department of Physics and ITCP, University of Crete, Voutes, GR-71003 Heraklion, Greece}
\affil{IESL and Institute of Astrophysics,
Foundation for Research and Technology-Hellas, PO Box 1527, GR-71110 Heraklion, Greece}

\author{Christopher Q. Trinh}
\affil{USRA/SOFIA, NASA Armstrong Flight Research Center, Building 703, Palmdale, CA 93550, USA}

\author{C. G. Volpert}
\affil{University of Chicago, Chicago, IL 60637, USA}

\author{Michael Werner}
\affil{NASA Jet Propulsion Laboratory, California Institute of Technology, 4800 Oak Grove Drive, Pasadena, CA 91109, USA}

\author[0000-0002-7567-4451]{Edward J. Wollack}
\affil{NASA Goddard Space Flight Center, Greenbelt, MD 20771, USA}

\collaboration{(HAWC+ Science Team)}

\begin{abstract}

We present Far-Infrared polarimetry observations of M82 at 53 and $154~\micron$ and NGC 253 at $89~\micron$, which were taken with HAWC+ in polarimetry mode on the Stratospheric Observatory for Infrared Astronomy (SOFIA). The polarization of M82 at $53~\micron$ clearly shows a magnetic field geometry perpendicular to the disk in the hot dust emission. For M82 the polarization at $154~\micron$ shows a combination of field geometry perpendicular to the disk in the nuclear region, but closer to parallel to the disk away from the nucleus. The fractional polarization at $53~\micron$ $(154~\micron)$  ranges from 7\% (3\%) off nucleus to 0.5\% (0.3\%) near the nucleus. A simple interpretation of the observations of M82 invokes a massive polar outflow, dragging the field along, from a region $\sim 700$~pc in diameter that has entrained some of the gas and dust, creating a vertical field geometry seen mostly in the hotter $(53~\micron)$ dust emission. This outflow sits within a larger disk with a more typical planar geometry that more strongly contributes to the cooler $(154~\micron)$ dust emission. For NGC 253, the polarization at $89~\micron$ is dominated by a planar geometry in the tilted disk, with weak indication of a vertical geometry above and below the plane from the nucleus. The polarization observations of NGC 253 at $53~\micron$ were of insufficient S/N for detailed analysis.

\end{abstract}



\section{Introduction} \label{sec:intro}

Starburst galaxies are an important phenomenon in the universe due to the presence of enhanced star formation and the accompanying strong outflows into the intergalactic medium. This type of galaxy might be an important contributor to the magnetization of the intergalactic medium in the early Universe \citep[e.g,][]{kron99, bert06}, but the generation and morphology of magnetic fields in starburst galaxies is poorly understood. 
Galactic scale winds are expected to be important at high redshift where starburst galaxies should be much more common than at the present epoch \citep[e.g.][] {veil05}. Nearby starburst galaxies with massive outflows provide an excellent laboratory for the study of starburst--driven winds where we can spatially resolve the wind and study the magnetic field geometry in detail. The relationship between spiral arms, outflows and galaxy-galaxy interactions and the magnetic field geometry has been extensively investigated in the radio \citep[see][for a review]{beck15, beck13}. Radio synchrotron emission arises from relativistic electrons and may not sample the same volume of gas as interstellar polarization, which is created by extinction or emission from asymmetric dust grains aligned with respect to the ambient magnetic field \citep[e.g.][]{jowh15}.  

An early suggestion that galaxies with strong infrared emission may be undergoing intense star formation was made by \cite{harp73}, based on Far-Infrared (FIR) observations of M82 and NGC 253. Because it is so well studied, M82 is considered the archtypical starburst galaxy \citep[e.g.][]{tele88, tele89} with an infrared luminosity of $3 \times 10^{10}$L$_ \odot$ and a star formation rate estimated at $13\rm{M_\odot}\;{\rm{y}}{{\rm{r}}^{{\rm{ - 1}}}}$. Based on extensive NIR integral field spectroscopy, \cite{fors03} find M82 is forming very massive stars ($\gtrsim$50-100 M$_\odot$). Their analysis suggests the global starburst activity in M82 occurred in two successive episodes. The first episode took place $10^7$ yr ago and was particularly intense at the nucleus, while the second episode occurred $5\times 10^6$ yr ago, predominantly in a circumnuclear ring and along what is believed to be a central stellar bar \citep[e.g.][]{lark94}. 

Similar to many galaxies with intense starbursts in their nuclear regions, M82 has a bipolar superwind emanating from the central region, stretching well into the outer halo area \citep[e.g.][]{shop98, ohym02, enge06}. The geometry of the magnetic field in the central starburst and in the superwind can be investigated using a number of different techniques. Classic interstellar polarization (in extinction) at $1.65~\micron$ in the Near-Infrared (NIR) by \cite{jone00} showed evidence for a near vertical field geometry at the nucleus. However, optical and NIR polarimetry is strongly contaminated by scattering of light from the very luminous nucleus of M82. 

Radio synchrotron observations also trace magnetic fields. \cite{reut94} found that the center of M82 is largely depolarized due to differential Faraday rotation. They find evidence for a vertical field in the northern halo and a more planar geometry in the southwestern disk region. \cite{adeb17} also find a planar geometry in this region and propose a magnetic bar that stretches across the entire central region. They detect some polarization at the nucleus and find the field is vertical there and can be traced out to 600pc (35\arcsec, assuming a distance of 3.6 Mpc \citep{kara06}) into the halo. They conclude that some of the non-detection of polarized emission 200pc North from the nucleus is most likely caused by canceling of polarization by the superposition of two perpendicular components of the magnetic field along the line of sight.

At a distance of 3.5 Mpc \citep{reko05}, NGC253 is also a well studied starburst galaxy with a strong galactic wind \citep[e.g.][]{shar10} and strong Mid-Infrared (MIR) and FIR emission \citep{riek80}. Unlike M82, which is seen nearly edge--on, NGC 253 has a visible tilt of about $12\degr$ \citep{deva58}, exposing its nuclear regions and revealing a spiral pattern in the disk. The outflow, projected against the tilted disk, is not as prominent as in M82. However, as in M82, it is seen in several tracers, including emission from dense molecular gas \citep{bolatto13, walt17}. Radio observations of NGC 253 do not show evidence for a vertical magnetic field geometry in the nucleus \citep{hees09}. Rather, the polarization map is consistent with a largely planar (disk) geometry.

Polarimetry at FIR wavelengths does not suffer from scattering effects. The albedo is $\gamma<10^{-5}$, \cite{drai84}. Faraday rotation, which is proportional to $\lambda^2$ \citep[eq. 3-71][]{spit78}, is also insignificant. Also, it traces the column density of dust, which is much more closely tied with the total gas column density than the relativistic electrons producing the radio synchrotron emission. While the FIR--Radio emission correlation \citep{helo85} might suggest that the energy in cosmic ray electrons and the thermal dust emission are strongly associated, it is not clear that they trace the same magnetic field geometry. With the advent of a FIR polarimetric imaging capability on SOFIA via HAWC+ \citep{harp18}, we can now map the magnetic field geometry in both the disk and central regions of M82 and NGC 253 with the goal of understanding the role of magnetic fields in starburst galaxies. 

\section{FAR--IR Polarimetric Observations} \label{sec:obs}

M82 and NGC 253 were observed as part of the Guaranteed Time Observation program with the High-resolution Airborne Wideband Camera-plus (HAWC+) \citep{vail07, dowe10, harp18} on the 2.5-m Stratospheric Observatory For Infrared Astronomy (SOFIA) telescope. We made observations of the inner regions of these galaxies using the standard chop-nod polarimetry mode with the instrumental configurations shown in Table~\ref{tbl:obs}.  HAWC+ polarimetric observations simultaneously measure two orthogonal components of linear polarization using a pair of detector arrays with 32 columns $\times$ 40 rows each.  Observations were acquired with a sequence of four dither positions in a square pattern with half side length of three pixels.  Integrations with four half-wave plate angles were taken at each dither position.  Based on the morphology of the sources evident in Herschel maps \citep{rous10, pere18}, the chop throw and angle are sufficient to make the intensity in the chop reference beams negligible for the results reported here.

Data were reduced using the HAWC+ Data Reduction Pipeline v1.3.0beta3 (April 2018), but with some customizations to address these particular data sets.  The data for all dither positions were screened for quality in the telescope tracking and basic instrument function, resulting in exclusion of one dither position for NGC 253 at 89 \micron .  As is standard with v1.3.0beta3, a $\chi^2$ test was performed by intercomparing dither sets.  We found that the statistical uncertainties were underestimated by a small, typical amount, and we inflated the uncertainties to account for this.  The inflation factors ranged from 1.19 to 1.34 for Stokes Q and U.


For subtraction of instrumental polarization, we used a revised calibration data product (v2, Aug. 2018); this is based on ``polarization skydips'' as is the previous standard calibration product and differs by only $\Delta q$ or $\Delta u \approx 0.1\%$ from it, but it is believed to be more accurate.

Close inspection of the M82 154 \micron\ data revealed a likely crosstalk effect in the detector system, in which a bright source produces a small, artificial response in another detector in the same readout column.  The magnitude of this effect is approximately 0.01-0.1\% of the signal in the detector with the bright source, and it seems to affect a fraction of the rows which are read out following the bright source.  This crosstalk was confirmed in separate observations of planets, and it may be similar to a crosstalk effect observed in certain calibration data from BICEP2 \citep{brevik12}, which has some cryogenic and room-temperature detector readout circuit designs in common with HAWC+.  We have not yet been able to develop a detailed model for the crosstalk, including its apparent variability over time; instead, we identify measurements (with granularity of ``dither position'' and ``detector pair'') which are likely to be affected significantly by the crosstalk and discard them.

Specifically, we looked at positions away from the bright cores of the galaxies for inconsistency in the ``$R+T$'' total intensity signal among the 8 nods comprising the fundamental polarization measurement sequence (a single dither position with 4 half-wave plate angles).  The $R+T$ detector pair sum \citep{harp18} should be constant over the 8 nods (independent of polarization) except for noise, calibration drifts, and artifacts in the detector system.

For the M82 154 \micron\ data, most of the suspect measurements identified and removed were found in the same column as bright galaxy emission and in rows which are read out afterward —- indicative of the crosstalk effect.  For the other data sets, we found fewer inconsistent $R+T$ measurements, and the majority of them correspond to otherwise noisy detectors.  We removed 0.03-0.7\% of the measurements with this first cut, depending on the observation.  We followed this with a general-purpose deglitcher which operates in the map domain, as described by \citet{chus18}.
Approximately 0.1-0.7\% of measurements were removed by the deglitcher.  For the two epochs of the M82 154 \micron\ observations, the parallactic angle differed by $\sim80\degr$, which improved the deglitching performance and map uniformity.

For M82 at both wavelengths and NGC 253 at 89 \micron , spatially-extended polarization was detected with statistical significance in excess of 10$\sigma$ in parts of each map.  Our 53 \micron\ observation of NGC 253 has significantly lower signal-to-noise in polarization, however, and since the effective integration time across the galaxy nucleus varies significantly due to the source falling on inoperational detectors for several of the dither positions, the polarization map is difficult to interpret.  For this observation, we report only an integrated signal in Table~\ref{tbl:integrated}.

\begin{deluxetable}{cccccccc}
\caption{Observation log.}
\tablewidth{0pt}
\tablehead{\colhead{Object} & \colhead{$\lambda_c$} & \colhead{FWHM$^a$} & \colhead{Obs. Date} & \colhead{Chop-Throw} & \colhead{Chop-Angle} & \colhead{DP$^b$} & \colhead{Obs. Time} \\
\colhead{} & \colhead{$\micron$} & \colhead{$\arcsec$} & \colhead{Y/M/D} & \colhead{$\arcsec$} & \colhead{$\degr$, E of N} & \colhead{} & \colhead{sec} }
\startdata
M82 & 53 & 5.5 & 2016/12/08 & 180 & cross el. & 16 & 4243 \\
 & 154 & 15.3 & 2017/10/24 & 180 & cross el. & 8 & 2141 \\
 &     &      & 2018/09/27 & 180 & cross el. & 8 & 2089 \\
NGC253 & 53 & 5.5 & 2017/10/19 & 300 & -40 & 6 & 1660 \\
 & 89 & 8.8 & 2017/10/19 & 300 & -40 & 7 & 2283 \\
\enddata
\tablenotetext{$a$}{The tabulated FWHM includes smoothing in the map generation.}
\tablenotetext{$b$}{Number of dither positions observed, with each consisting of a complete half-wave plate cycle.}
\label{tbl:obs}
\end{deluxetable}

\section{M82: Dust Temperature and Column Density} \label{sec:temp}

\begin{figure} 
\begin{center}
\includegraphics[width=2.9in]{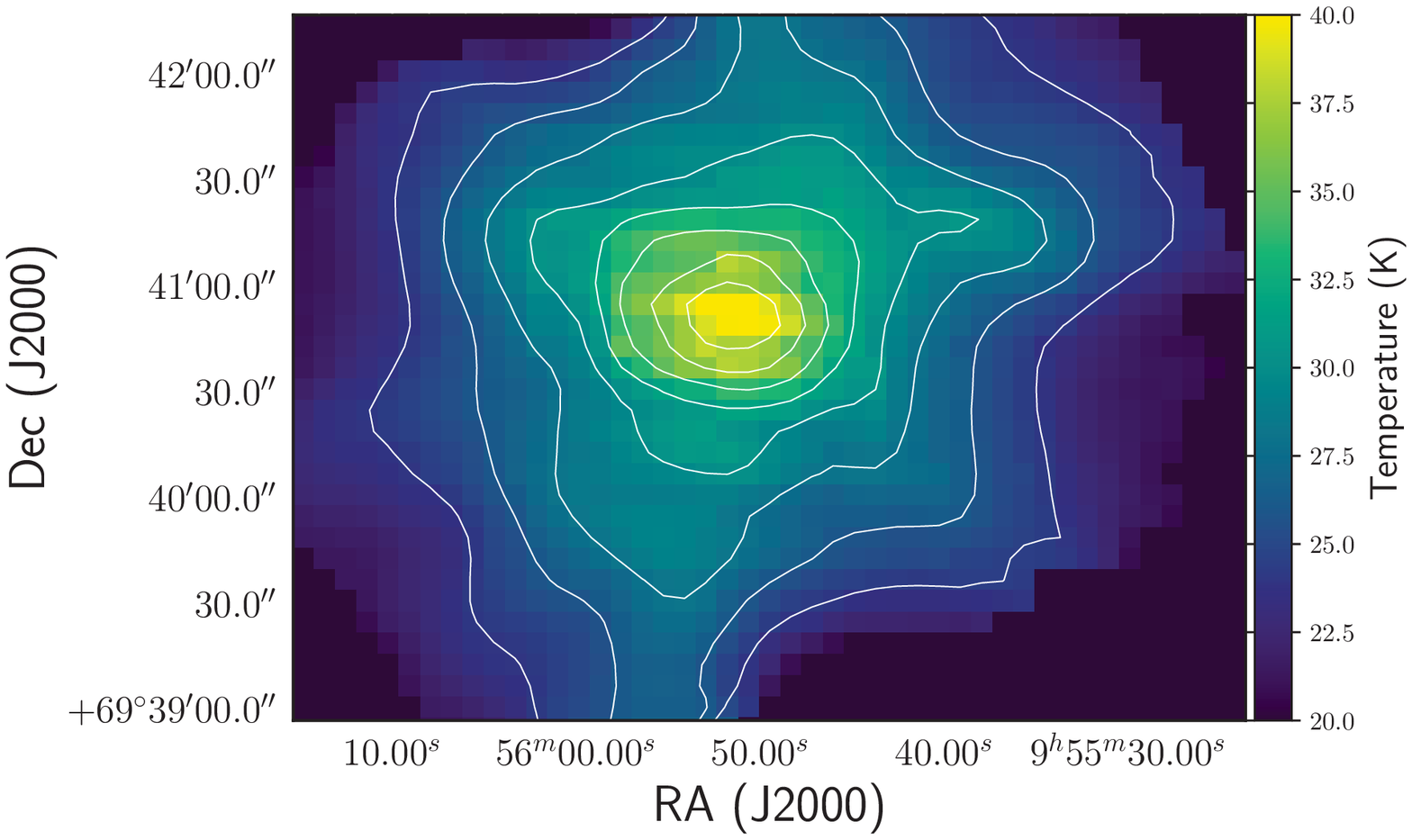}
\includegraphics[width=2.9in]{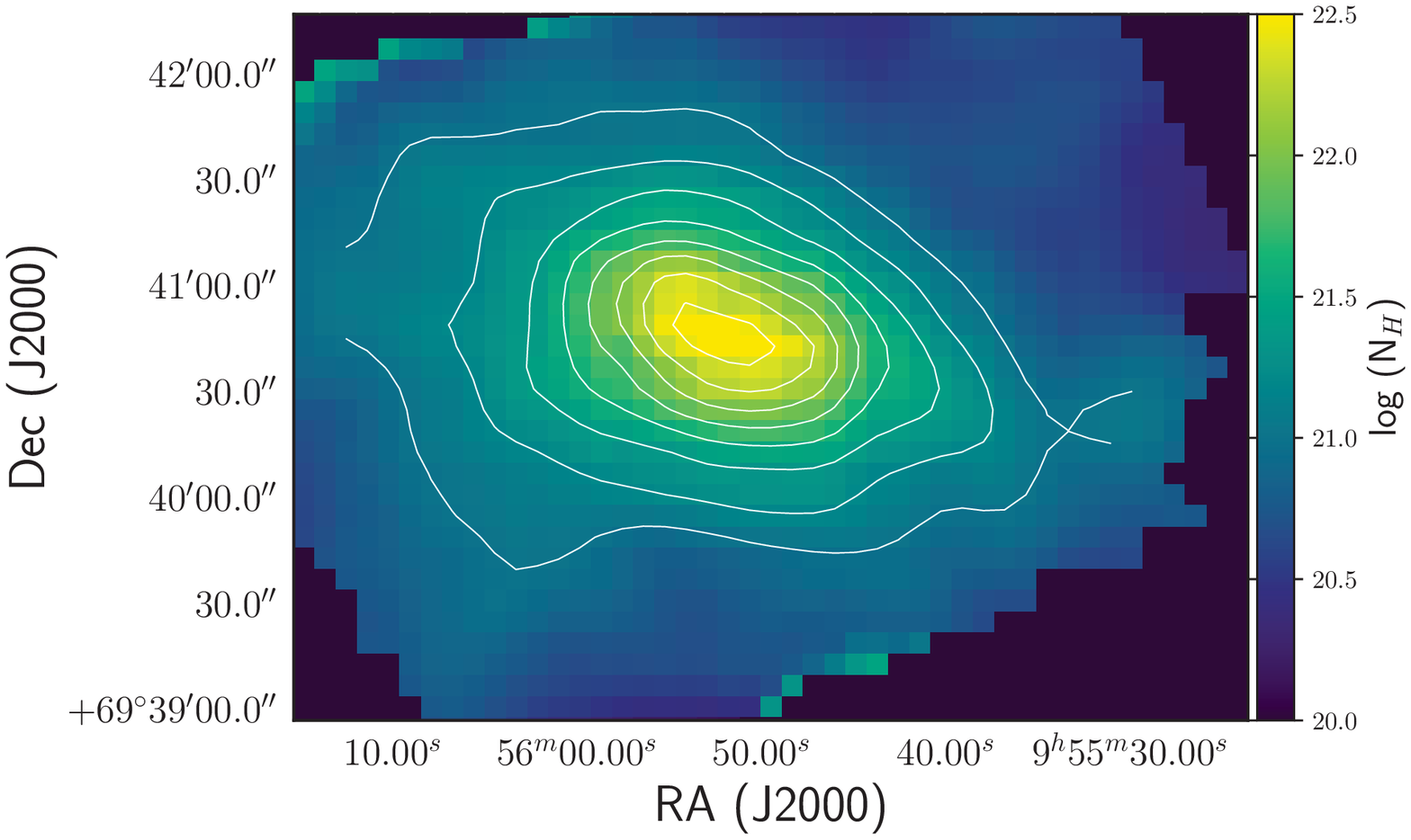}
\caption{M82 maps of color temperature (left) and column density (right). The temperature contours start at 20 K and increase in 2K increments. The column density contours start at $\log \left( {{N(H+H_2)}} \right) = 21$ and increase in increments of 0.2.}
\end{center}
\end{figure}

Previous lower spatial resolution M82 observations at FIR wavelengths by \cite{kane10} find a dust mass of $2.3 \times {10^6}~{{\rm{M}}_ \odot }$ within the central $2\arcmin$ using a multitiple--dust temperature fit to their AKARI data. Using a gas-to-dust ratio of 100, this corresponds to a total mass (gas and dust) $M_{tot} \sim 2 \times 10^8~\rm{M_\odot}$. By subtracting the FIR emission from the starburst and the disk of M82, \cite{rous10} estimate the total dust mass in the wind and halo from their $500~\micron$ and dust temperature maps to be $1.1 \pm 0.5 \times 10^6 ~\rm{M}_\odot$, or $M_{tot} \sim 1 \times 10^8~\rm{M_\odot}$. Using FORCAST on SOFIA, \cite{niko12} were able to image the inner $75\arcsec$ of M82 at wavelengths from $6.6 - 37.1~\micron$ at an angular resolution $\sim 4\arcsec$ (70pc). Their analysis uses measurements of extinction and surface brightness to constrain the total mass and dust mass of the two central peaks seen at NIR and MIR wavelengths. They find $\rm{A_V}=18$ and $\rm{A_V}=9$ toward the main and secondary emission peaks with an estimated color temperature of 68 K at both peaks. The dust masses at the peaks within a $6\farcs6 \times 6\farcs6$ region were estimated to be $\sim 10^4 ~\rm{M_\odot}$, or $M_{tot} \sim 10^6~\rm{M_\odot}$. Based on the M82 rotation curve \citep{grec12}, the total mass within a 100pc $(5.7\arcsec)$ radius of the center is $M_{100} \sim 5\times 10^7 \rm{M_\odot}$ and within 500pc $(28.6\arcsec)$ radius, $M_{500} \sim 1.5\times 10^9 \rm{M_\odot}$. 


To support the polarimetric analysis, we made temperature and column density maps of M82. Specifically, we combine our 53 and $154~\micron$ HAWC+ observations with the publicly available 70, 160, and $250~\micron$ observations from the Herschel Space Observatory \citep{pilb2010} with the PACS instrument (Poglitsch et al. 2010) and SPIRE instrument \citep{grif2010}. These observations were taken as part of the Very Nearby Galaxies Survey (PI: Christine Wilson).  We bin each observation to the pixel scale, $6\arcsec$, of the $250~\micron$ {\textit{Herschel}} image. We then extract the intensity values of each pixel associated to the same part of the sky at each wavelength. Finally, we fit a modified blackbody function assuming a dust emissivity of $\epsilon_\lambda \propto \lambda^{-1.6}$ \citep[e.g.][]{bose12}, with the temperature and optical depth at $250~\micron$, $\tau_{250}$, left as free parameters. We compute ${N(H+H_2)} = {\tau _{250}}/\left( {k\mu {{\rm{m}}_{\rm{H}}}} \right)$ where $k=0.1$cm$^2/g$ \citep{hild83} and $\mu$ = 2.8. We use the HAWC+ data to both augment the Herschel data and help constrain the Wien side of the SED at $53~\micron$. Figure 1 shows both the temperature and column density maps within the same FOV.

The color temperature ranges from a peak of 40 K on the nucleus to 25 K at about $20\arcsec$ along the disk to the NE and SW. The computed column density peaks at $N(H+H_2) = 3 \times 10^{22}~ \rm{cm^{-2}}$, about $\rm{A_V} \sim 20$, somewhat higher than found by \cite{niko12}. Summing the column density in a $40\arcsec \times 20\arcsec$ box yields $M_{tot} \sim 8 \times 10^7~\rm{M_\odot}$. This is at least a factor of two less than seen in the molecular gas in the same region. Given the filling factor of 1\% for the dense gas derived by \cite{nayl10}, most of the molecular gas, even if the temperature of the dust on the surface of the dense cores was as high as 50 K, is probably not contributing significant flux to our HAWC+ maps. This means our HAWC+ observations do not sample regions of very dense, molecular cores, but rather, they sample the dust associated with the rest of the ISM in M82, including less dense $(\rho \la 100~\rm{cm^{-3}})$ molecular gas.

\section{M82: Polarization Maps} \label{sec:maps}

The SOFIA observations of M82 are shown in Figure 2, where we have plotted polarization vectors on a grid with one half beam width for the spacing and with position angles rotated $90\degr$ to represent the inferred magnetic field direction. In the top row, the vector length is proportional to the fractional polarization. Cuts for the fractional polarization were made at a S/N of 3.3/1 (debiased, see \cite{ward74}) and at an intensity of 0.21 Jy/$\square\arcsec$ at $53~\micron$ and 0.044 Jy/\sq\arcsec\ at $154~ \micron$. Since HAWC+ is a relatively new instrument, we chose to be conservative in our S/N cuts. Also, since there is a large number of pixels outside these intensity contours $(\sim 1000)$, we used the cut in intensity to remove a few statistically insignificant pixels with no corroborating nearby position angles. We can likely trust the remaining vectors as being indicative of the field direction. In the second row, all vectors within these criteria are plotted with the same length to better clarify the position angle morphology. At a wavelength of $53~\micron$, the polarization fraction ranges from a high of 7\% well off the nucleus to values as low as 0.5\% at some locations along the plane (disk) of M82. The fractional polarization at the intensity peak is 2.2\% , and it declines toward the east and west along the plane.  Although the nominal systematic error in polarization for HAWC+ is 0.3\% (1 $\sigma$) \citep{harp18}, in this specific map, detections with $p$ as low as 0.5\% appear to have position angles which fit the large scale pattern. Line Integral Contour maps using lower S/N data are shown in Figure 3 to better illustrate the mean position angle at greater distances from the disk.  

Clearly evident in the $53~\micron$ image is the presence of a magnetic field geometry largely perpendicular to the plane. This geometry extends over a region at least 700pc ($40\arcsec$) along the disk and up to 350 pc above and below the plane. The stellar scale height for the thin disk in M82 is $h_z = 143$pc \citep{lim13}, however the more extended distribution of AGB stars is much greater \citep{davi08}. We do not measure the magnetic field geometry in the outflow at kpc scales.

\begin{figure}
\begin{center}
\includegraphics[width=2.9in]{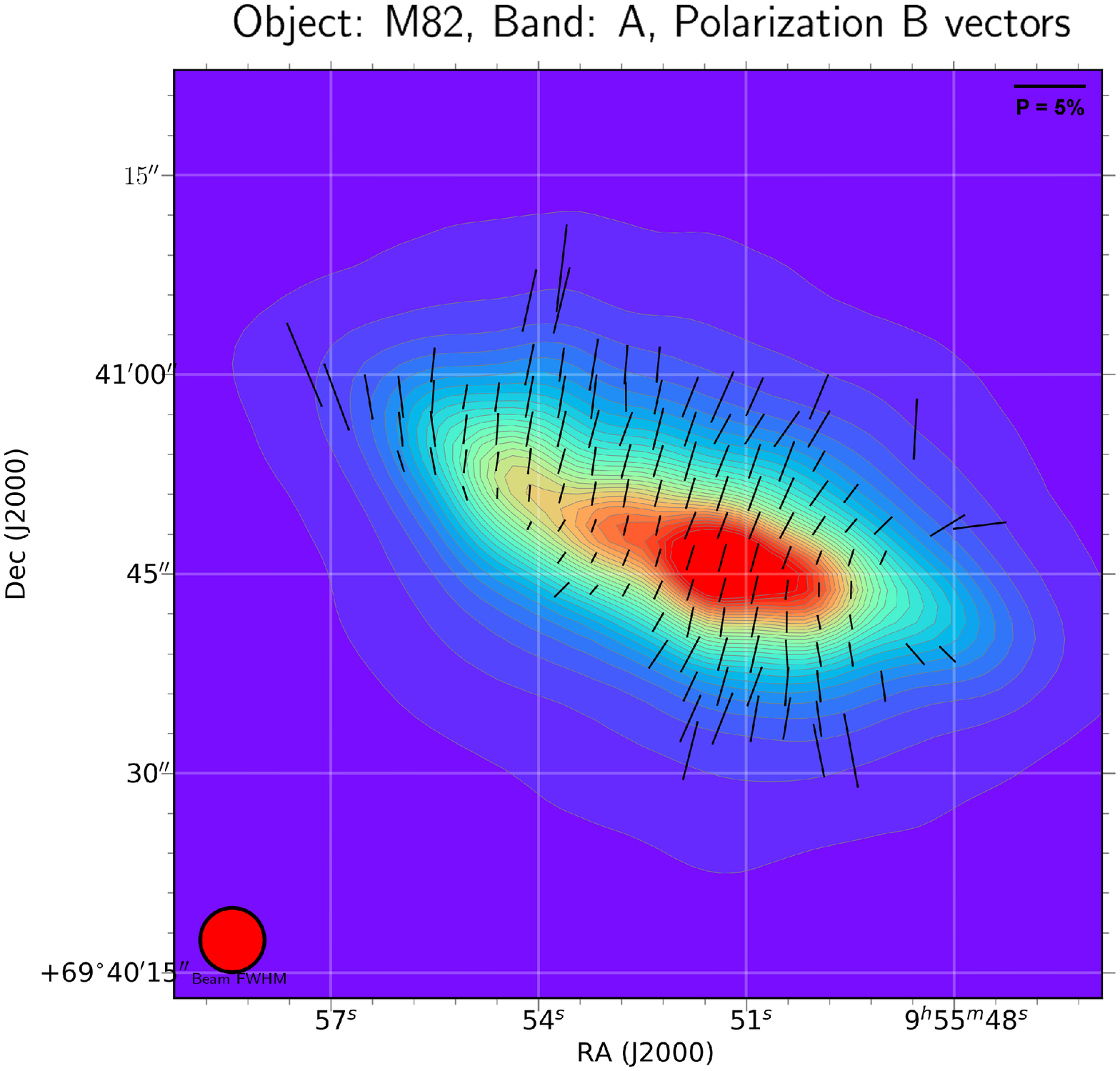}
\includegraphics[width=2.9in]{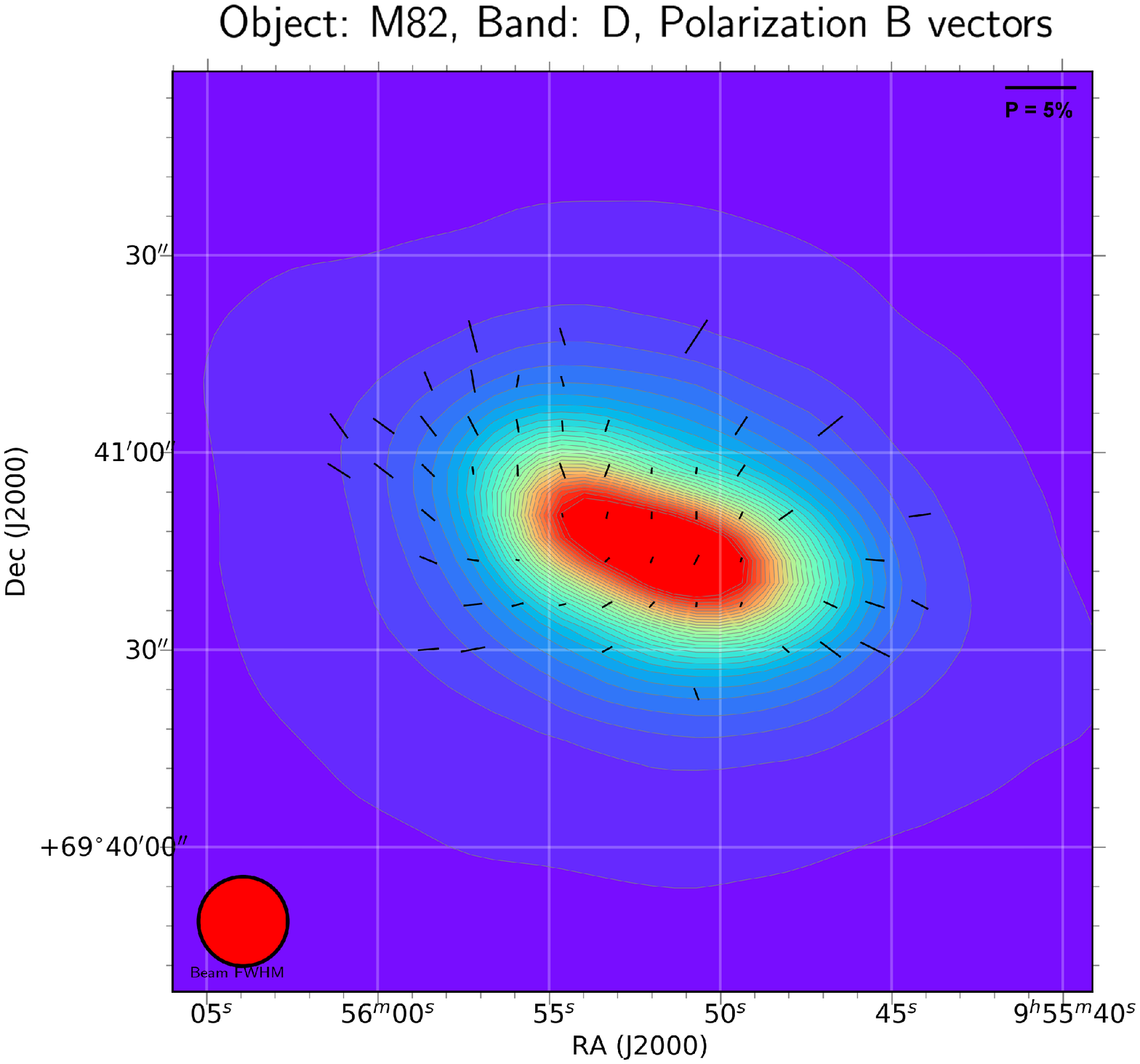}
\includegraphics[width=2.9in]{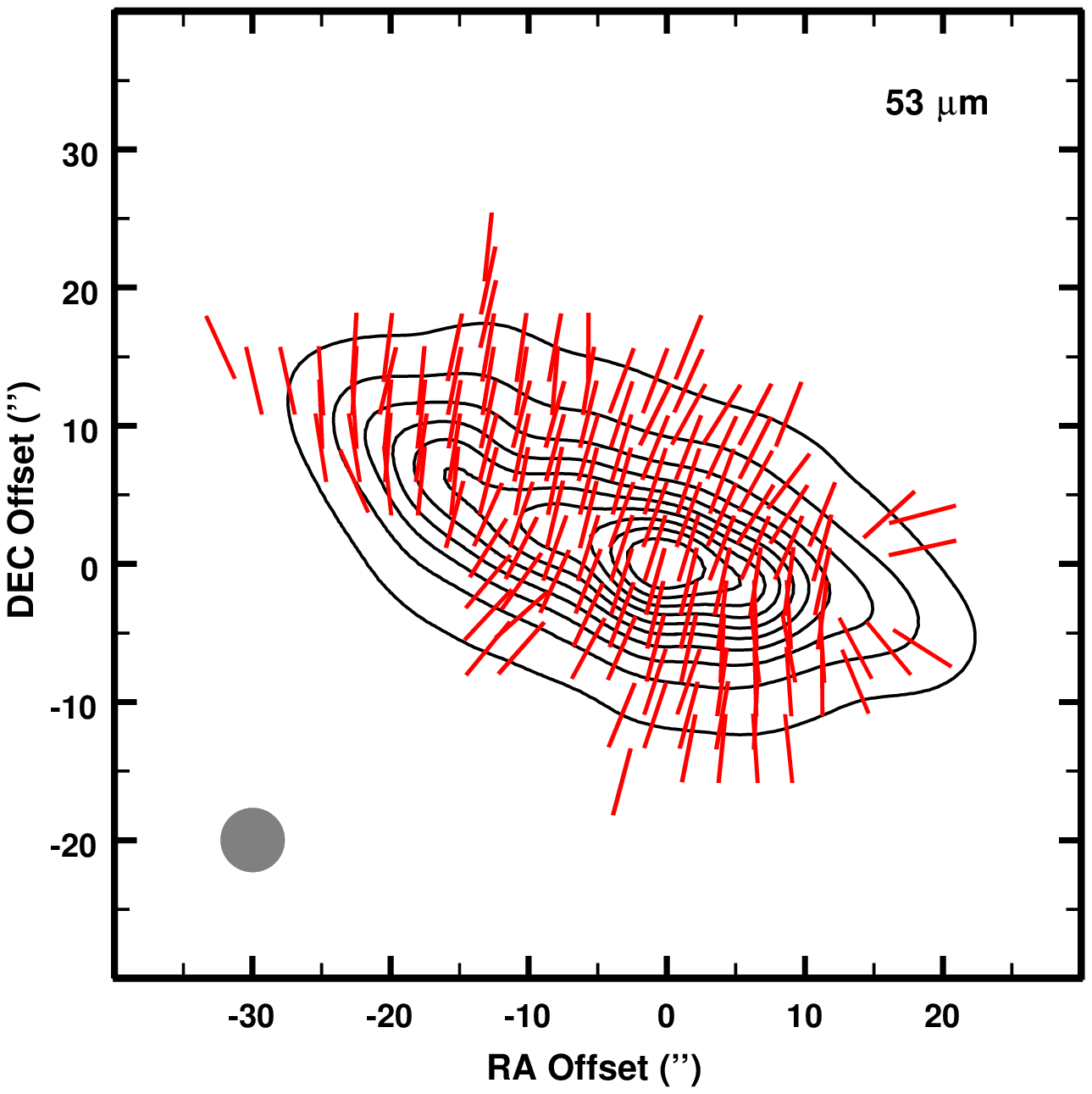}
\includegraphics[width=2.9in]{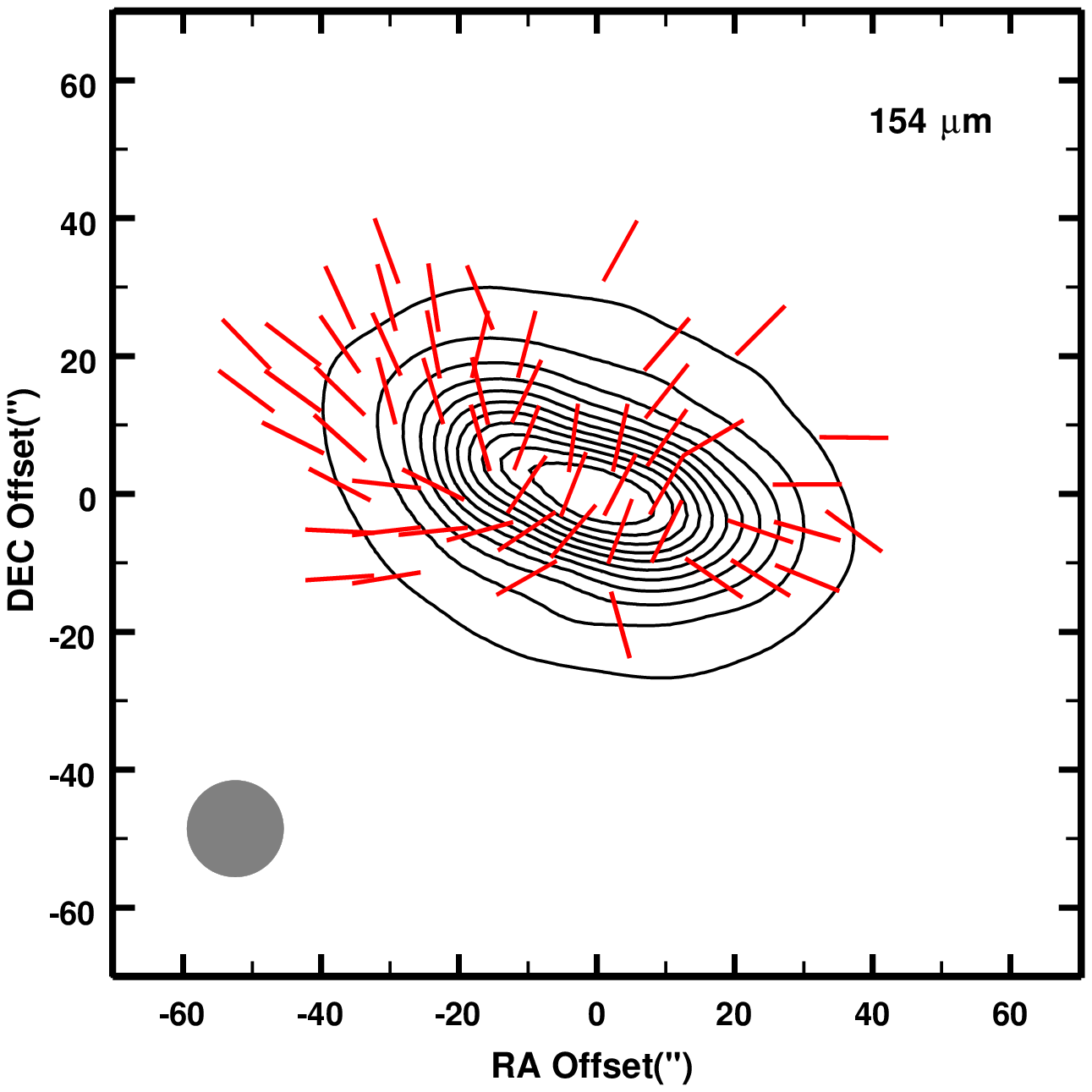}
\caption{Polarization vector maps of M82, rotated $90\degr$ to represent the inferred magnetic field direction. Upper left: $53~\micron$ fractional polarization.  Upper right: $154~\micron$ fractional polarization. Bottom left: Position angle only overlaying intensity contours. The first contour starts at $2\times 10^4$ MJy/sr with increments of $2\times 10^4$ MJy/sr. Bottom right: Position angle only overlaying intensity contours. The first contour starts at $4\times 10^3$ MJy/sr with increments of $4\times 10^3$ MJy/sr.}
\end{center}
\end{figure}

\begin{figure}
\begin{center}
\includegraphics[width=2.9in]{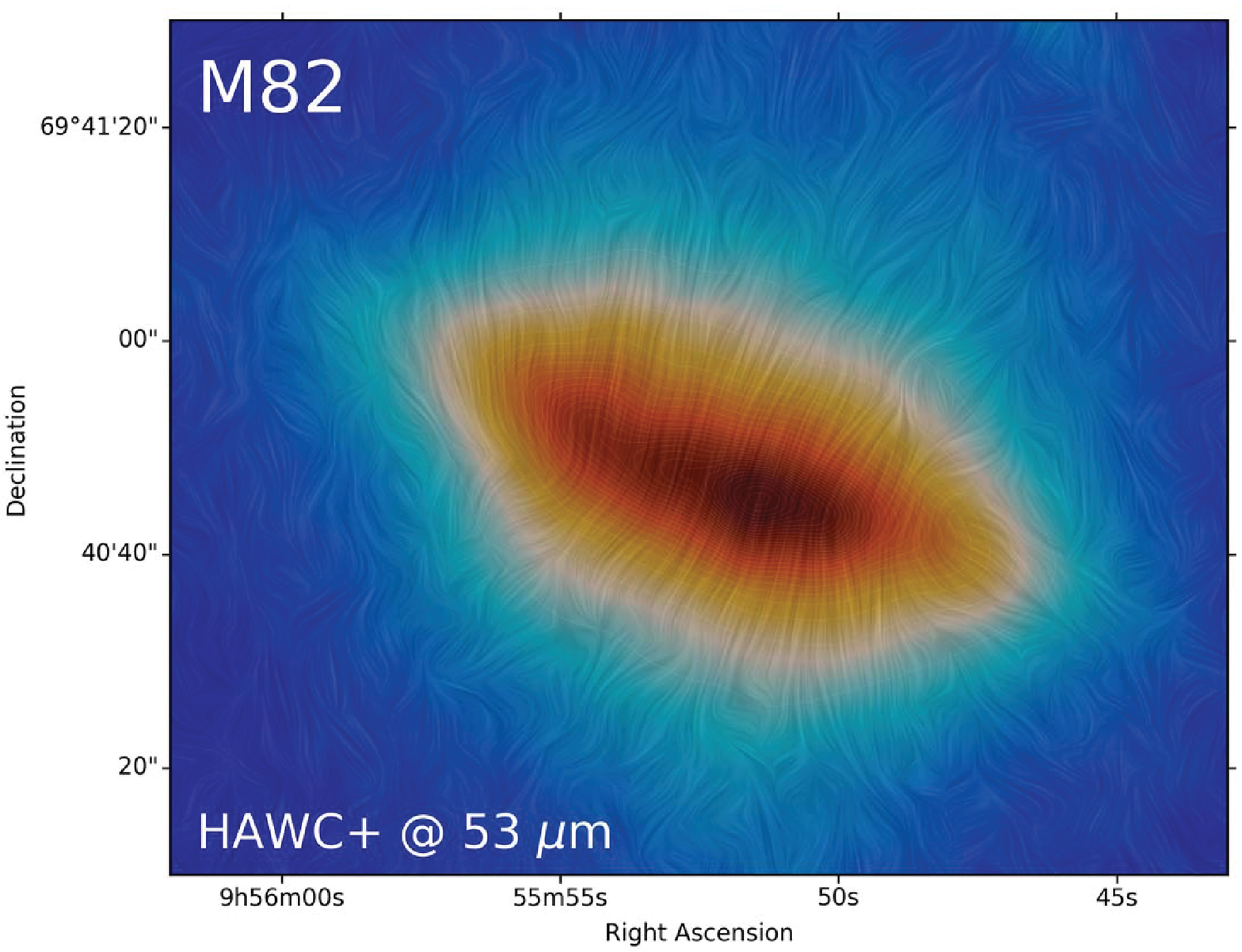}
\includegraphics[width=2.9in]{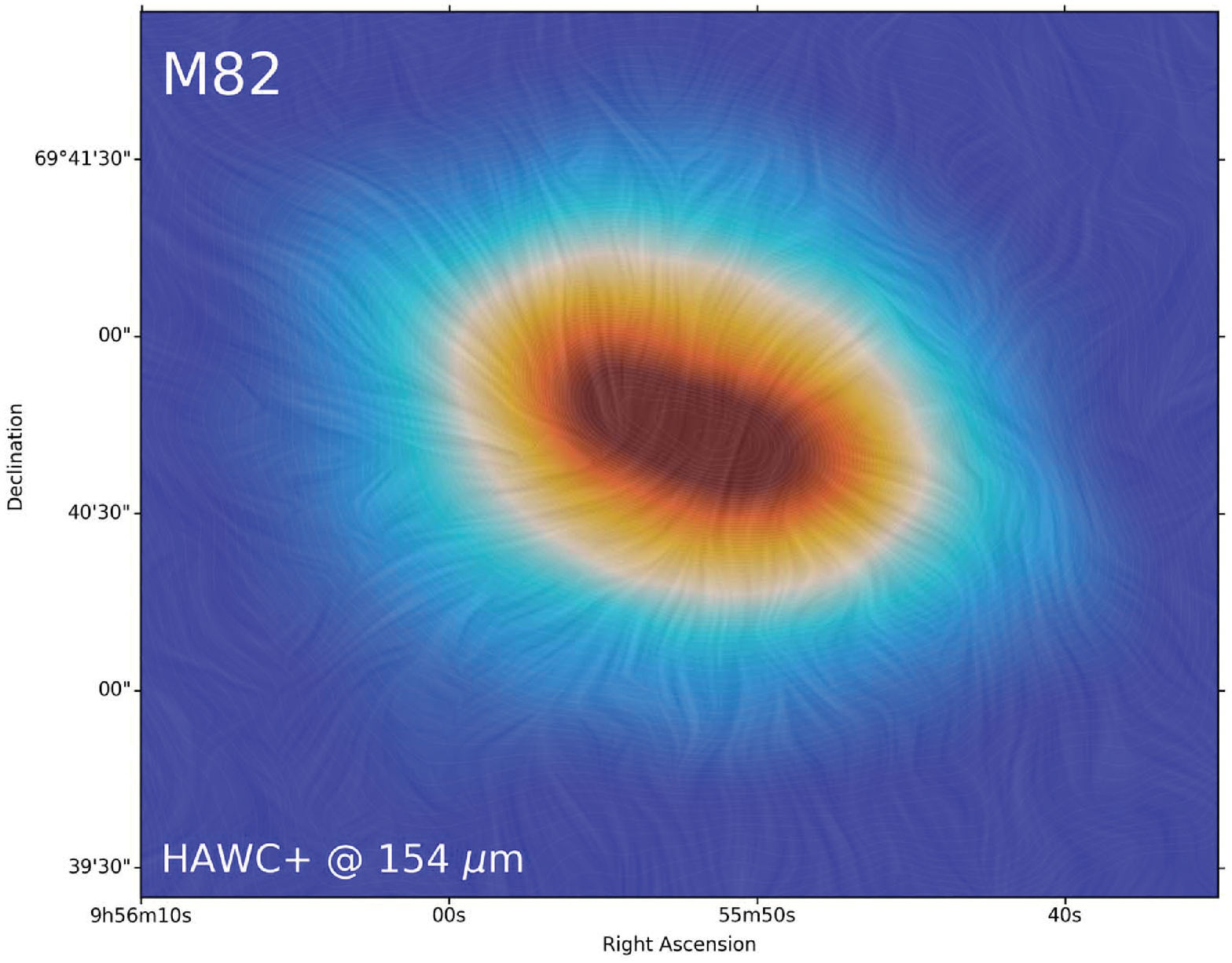}
\caption{Line Integral Contour \citep{cabr93} maps of the polarization data. Notice the transition of the position angle from vertical to planar to the SW in the $53~\micron$ map and to both the NE and SE in the $154~\micron$ map. A cut in polarization S/N (debiased) of $2\sigma$ was used to form these images, which allows the general trend in position angle to be traced further into the halo and along the disk.}
\end{center}
\end{figure}

For the polarization map at $154~\micron$, the polarization fraction ranges from a high of 3\% well off the nucleus to 0.3\% near the nucleus, and the vectors show more variation than in the $53~\micron$ map. The polarization vectors in the central region and to the North and Northwest are consistent with a vertical field.  The vertical field is displayed by vectors with fractional polarization ranging from 4\% in the Northwest to only 0.3\% in the disk, so some caution is needed in interpreting the detailed field structure.  The vectors to the Southwest and Northeast along the disk have larger magnitude (most $>$1\%) and indicate a geometry closer to parallel to the plane of the disk. Using NIR polarimetry in extinction, \cite{jone00} found evidence for a vertical geometry at the nucleus, but a planar geometry to the SW of the central region. Our FIR observations agree with the geometry found by \cite{jone00}, but the NIR observations were heavily contaminated by scattering polarization and were not conclusive. Our $154~\micron$ polarimetry is consistent with a planar disk geometry for the magnetic field visible in the polarization vectors to either side of the nuclear starburst. There is evidence for this in the $53~\micron$ map as well in a few vectors to the SW and perhaps the NE. 

Using the SCUBA camera on the JCMT, \citet{grea00} reported an 850 \micron\ polarization map for M82 consisting of 22 polarization vectors with $>3\sigma$ significance and covering a $40\arcsec \times 50\arcsec$ region similar to ours.  The main features of the map are a vertical magnetic field at the west nucleus, (inferred) low polarization at the east nucleus, and a loop or bubble shape to the field at the outskirts of the map.  \citet{matt09} reprocessed the $850~\micron$ data as part of an archive paper and produced a map with 16 vectors at $>2\sigma$ signficance.  The vertical field at the west nucleus is still apparent, as is the (inferred) low polarization at the east nucleus, but the loop is no longer clear.  The dust emitting at 850 \micron\ should also contribute signficantly in the HAWC+ $154~\micron$ band.  Comparing the maps at those two wavelengths, we find agreement regarding the vertical magnetic field toward the west nucleus and the decrease in fractional polarization toward the east.  However, we do not observe the loop field, nor do we see other clear similarities.  We performed a statistical test of agreement between each version of the 850 \micron\ map and our $154~\micron$ map.  For each reported 850 \micron\ vector, we interpolated the $154~\micron$ Stokes parameters $(Q_{154},U_{154})$ and formed the dot-product-like quantity $S = Q_{154}Q_{850} + U_{154}U_{850}$.  Positive values of $S$ indicate agreement of polarization angles within 45\degr , and negative values indicate disagreement greater than 45\degr .  We observe a positive correlation for the 2-3 850 \micron\ measurements toward the west nucleus, but for the remaining 13-20 measurements we find just as many positive as negative values of $S$.  Of all the far-infrared/submillimeter polarization maps discussed in this section, the HAWC+ 53 and $154~\micron$ maps are the only ones that clearly show a correlation over an extended area.

\cite{jone00}, observing polarization in extinction at $1.65~\micron$, interpreted the combination of a vertical position angle in the nucleus and a planar geometry in the disk as a mixture of two magnetic field geometries along the line of sight. The field in the central regions with the starburst is vertical, while the field in the surrounding disk is planar. In extinction, this causes a partial cancellation of the fractional polarization as the light from the nucleus first traverses the region with a vertical field and then traverses the disk with a planar magnetic field. In essence, the two regions act as crossed--polaroids (more accurately crossed-fields). Based on the expected level of fractional polarization for the measured extinction to the nucleus of M82 in the NIR \citep{jone89, jone93}, \cite{jone00} estimated that 2/3 of the dust along the line of sight to the nucleus has a vertical field, with the remainder of the dust lying in the disk passing in front of the nucleus. 

The polarization in extinction is a function of the column depth of dust along the line of sight, but not the dust temperature. In emission at FIR wavelengths, hotter dust will radiate more effectively at shorter wavelengths than cooler dust. Referring to our dust temperature map, the $53~\micron$ emission is more sensitive to temperature than the $154~\micron$ emission because it is on the Wien side of the SED. Hence, the $53~\micron$ emission can dominate over the $154~\micron$ emission for regions along a path with warmer dust. Since the dust in the central region is hotter than in the disk (see Figure 1), the transition from vertical to planar position angle will take place more quickly at $154~\micron$ than at $53~\micron$ as the line of sight moves away from the nucleus along the disk. This is what we observe. If the dust temperature were constant everwhere in the disk, then the transition of the magnetic field geometry from planar to vertical would presumably take place at the same location for both wavelengths. 

The vertical magnetic field geometry we see in the HAWC+ FIR polarimetry lies along the same direction seen in other measurements of the super--galactic wind in M82. Optical and H$\alpha$ images suggest a conical outflow with a fairly narrow launch point in the nucleus. Line splitting seen in CO observations of the molecular gas is interpreted as due to a conical outflow perpendicular to the disk with an opening angle of $20\degr$ that stretches up to 1.5kpc from the nucleus \citep{walt02}. \cite{mart18} show that the HI kinematics are inconsistent with a simple conical outflow centered on the nucleus, but instead require the more widespread launch of the HI over the $\sim 1$kpc extent of the starburst region. This result is consistent with our finding that the region in the disk with a vertical magnetic field is at least 700pc wide. There is some evidence that the polarization vectors in the $154~\micron$ map to the East line up with streamers S2, S4 and possibly S3 seen in the CO observations of \cite{walt02}.

\section{M82: The Threaded Field} \label{sec:thread}

The magnetic field in the ISM of spiral galaxies has both constant (threaded) and turbulent components \citep[see a recent treatment by][]{plan18}. The effects of a turbulent component can be seen in both variations of the polarization position angle with position on the sky using a type of structure function \citep{kobu94, hild99, plan18} and the trend of fractional polarization with column depth \citep[e.g.][]{joba15, hild99}. We will examine the structure function in a later paper, but we can easily examine the trend of fractional polarization with optical depth. If the magnetic field geometry is perfectly constant with no bends or wiggles, the fractional polarization in emission will be constant \citep[][for a review]{joba15, jowh15} with optical depth in the optically thin regime. If there is a region along the line of sight that has completely unaligned grains, it will add total intensity to the beam, but no polarized intensity, resulting in a slope of $P \propto \tau ^{ - 1}$ with increasing contribution from that region \citep{joba15}.

If the dust grain alignment angle varies in a purely stochastic way along a line of sight, the fractional polarization in emission will decrease as $P\propto \tau^{-1/2}$ \citep{joba15}, and there will be no correlation in position angle across the sky. A combination of a constant and a purely random component will cause the polarization to decrease with optical depth at a rate inbetween these two extremes, with the constant component dominating the position angle geometry after several decorrelation lengths \citep{zwei96}. If there is a coherent departure from purely constant component such as a spiral twist, paths with perpendicular magnetic fields, or other smooth variations of the projected field along the line of sight that depend on total column depth, the fractional polarization can drop \textbf{faster} than $P\propto \tau^{-1/2}$ due to strong cancellation of the polarization. Note, in this context, we are considering the field in the outflow of M82 to have a threaded component aligned with the outflow.

We are assuming the efficiency of the grain alignment mechanism is not a factor in our FIR polarimetry of M82 \citep[see][for a review of grain alignment]{ande15}. We argued in \S \ref{sec:temp} that our observations are sensitive to the warm dust in the diffuse ISM of M82, but are largely insensitive to contributions from very dense, molecular cloud cores. There is good evidence that grain alignment in the Milky Way is at its maximum in the diffuse ISM, and only in dense cloud cores with no internal radiation field is there a possible loss of grain alignment \citep{joba15}. Since our HAWC+ observations are not sensitive to very dense molecular cloud cores, we do not expect regions with unaligned grains to contribute to our FIR polarimetry. We can not rule out that regions of very high turbulence (scrambled field) on small scales may be present along some lines of sight, mimicking regions with unaligned grains (adding net total intensity, but no net polarized intensity).

Figure 4 plots the trend in fractional polarization with column density at $53~\micron$ and $154~\micron$. The $3.3 \sigma$ upper limits are plotted as green triangles. Simple power--law fits to the upper bound of the data points in Figure 4 are steeper than $P\propto \tau^{-1/2}$ at $53~\micron$ and about the same at $154~\micron$. We are concentrating on the upper bound in these plots because that delineates lines of sight where the minimum depolarization effects are present. Lines of sight with lower $P$ could suffer significant depolarization effects, but it is hard to point to a specific line of sight and conclude which effects are dominant. If a `crossed--field' effect is at work along lines of sight through the plane of the disk, and it spans most of the area we have mapped, then it will lower the overall fractional polarization. Although the slope of $P \rm{vs.} N(H+H_2)$ at $53~\micron$ is about $P\propto \tau^{-1}$, the clear coherence of the position angles across the face of M82 indicates that a systematic cancellation of polarization is most likely taking place. 

The fraction of the gas with a vertical field is difficult to determine. If the very low fractional polarization in the nucleus is due to simple cancellation of polarization by the superposition of a planar disk field and a polar nuclear field, then (using the value of 2/3 for the column depth corresponding to the vertical component in \cite{jone00}) approximately $5-6 \times 10^7~\rm{M_\odot}$ has entrained a vertical field. If, however, turbulence on a small scale relative to our beam dominates the field geometry in the warm dust, leaving only a modest fraction of the dust with a coherent polar field, then this must be considered an upper limit. If the estimates of the molecular gas mass in M82 are correct (see \S \ref{sec:temp}), then a sigificant fraction of the mass in the nuclear region of M82 is not detected by our FIR polarimetry.

\begin{figure}
\begin{center}
\includegraphics[width=2.9in]{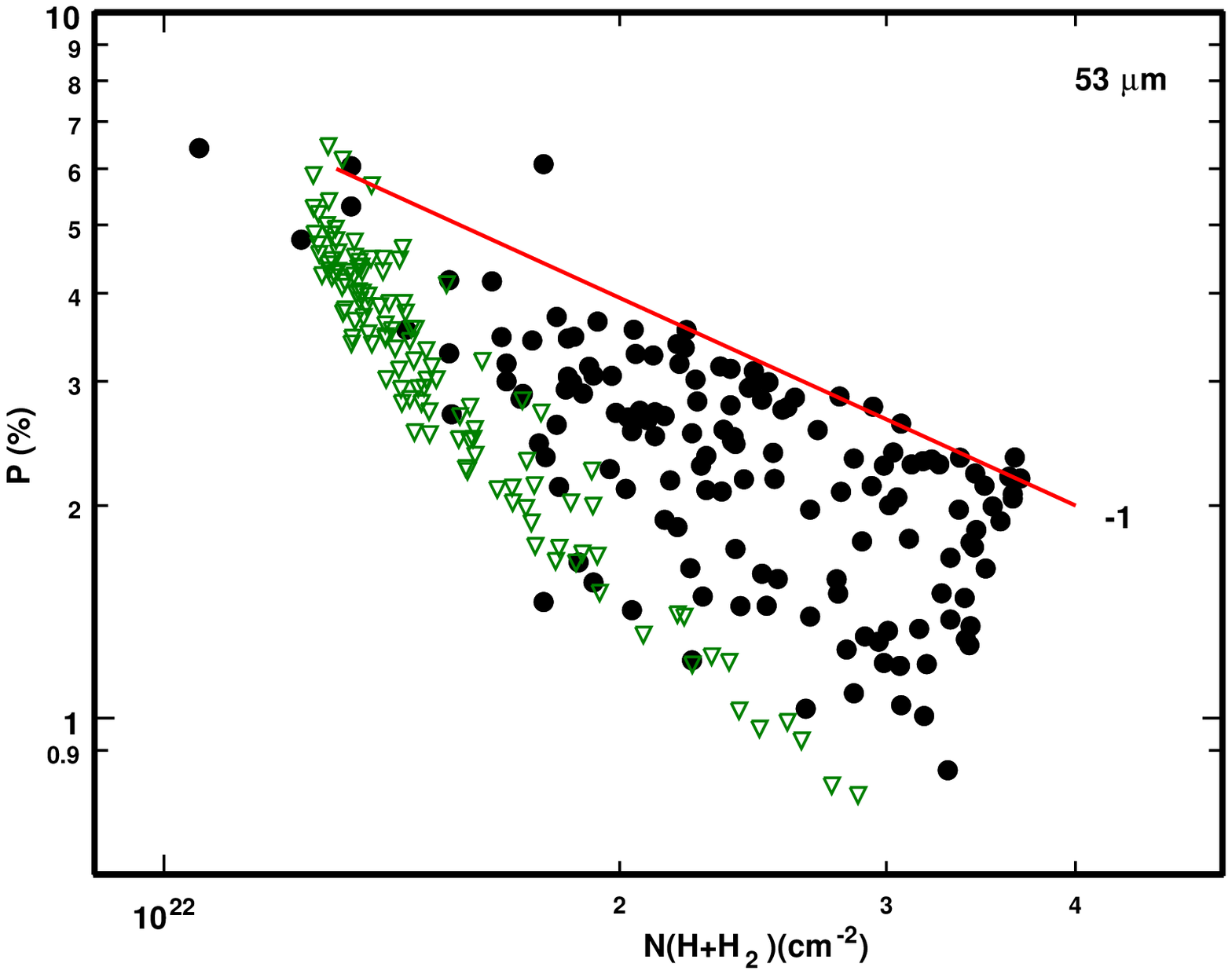}
\includegraphics[width=2.9in]{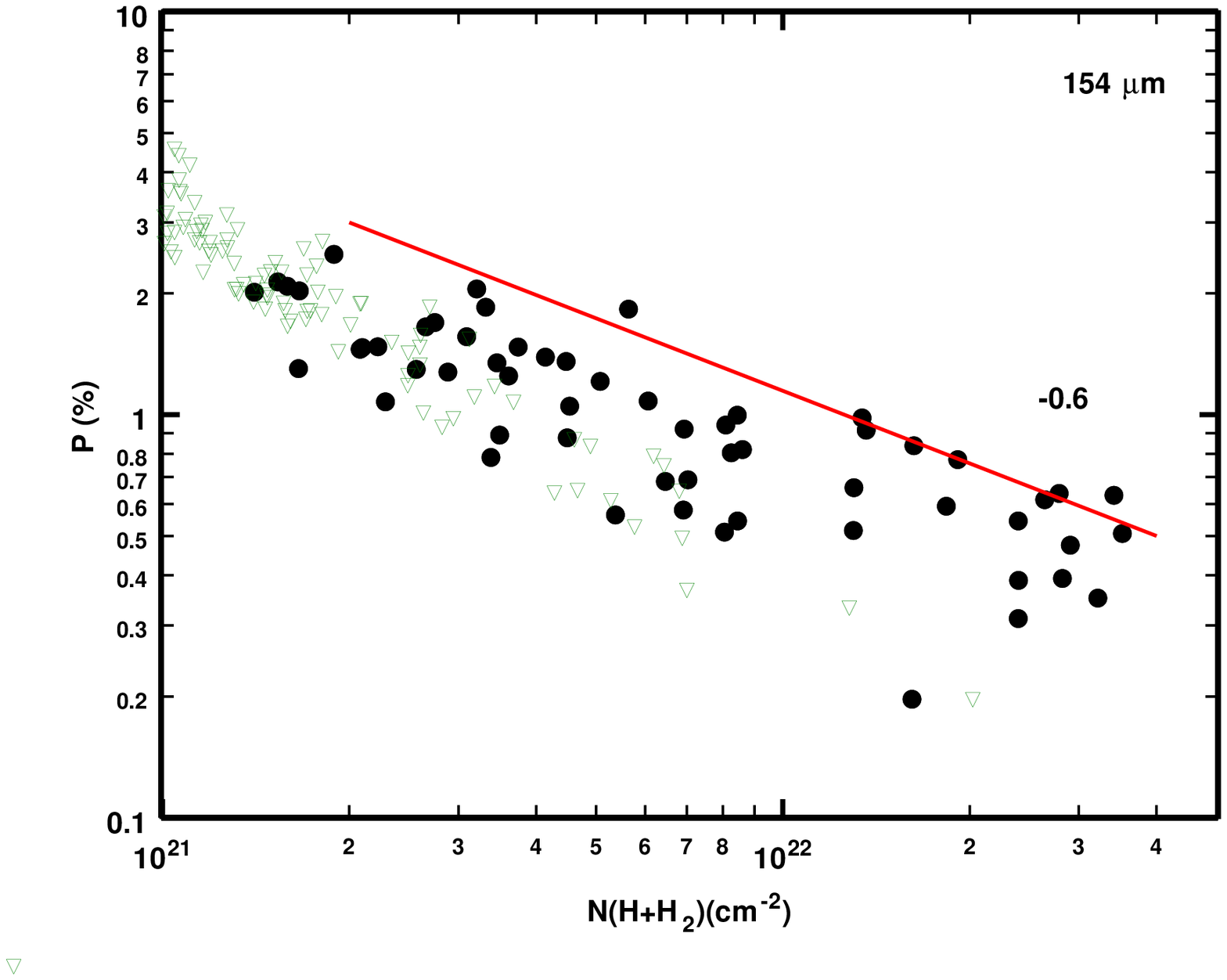}
\caption{Plot of fractional polarization against $N(H+H_2)$ column depth for high S/N data points at $53~\micron$ and $154~\micron$ for M82. The $3.3 \sigma$ upper limits are plotted as green triangles.}
\end{center}
\end{figure}

\section{NGC253}

In this section, we present the $89~\micron$ results of NGC 253, and these observations have an interesting contrast to those of M82.

\begin{figure}
\begin{center}
\includegraphics[width=2.9in]{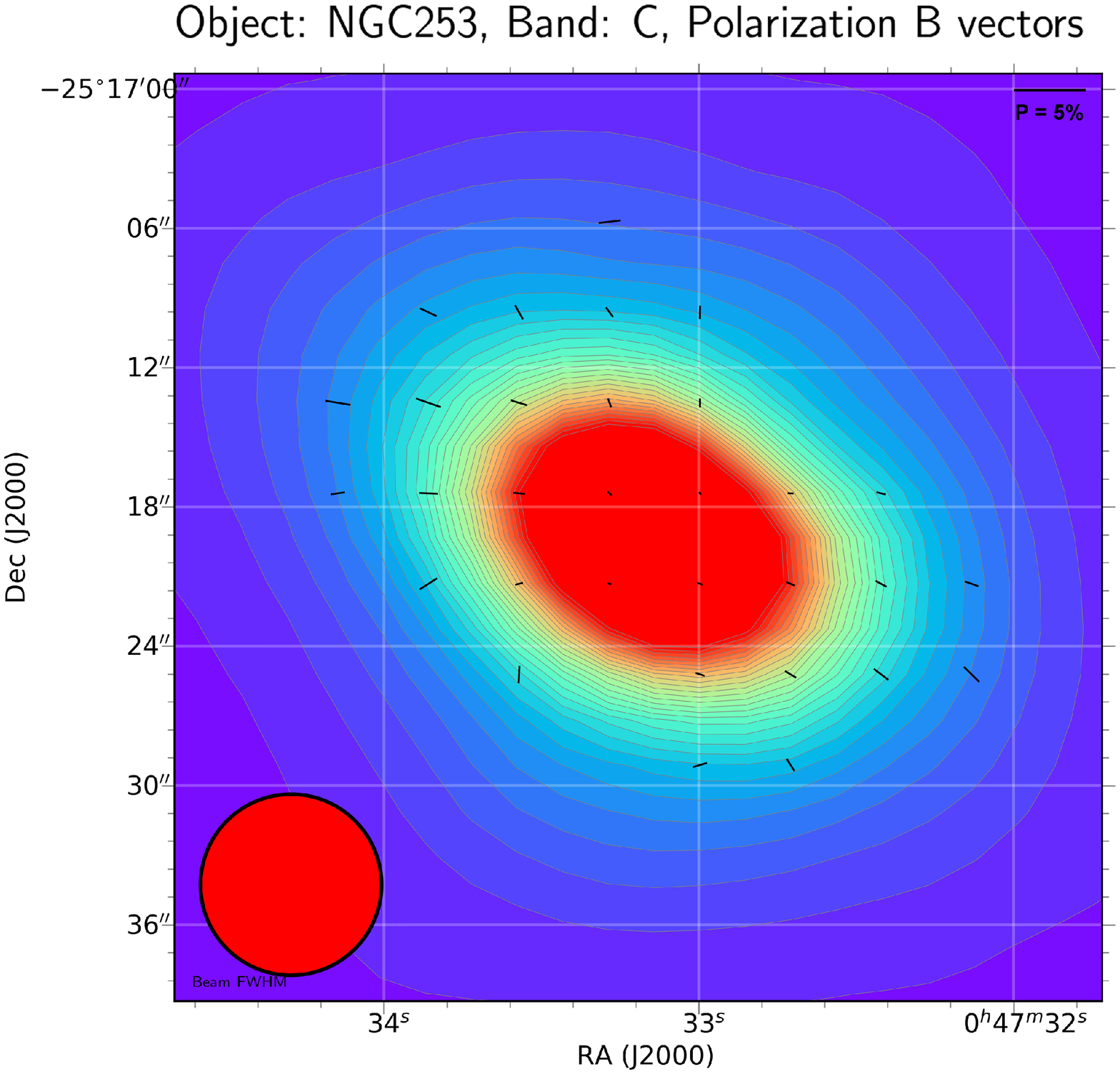}
\includegraphics[width=2.9in]{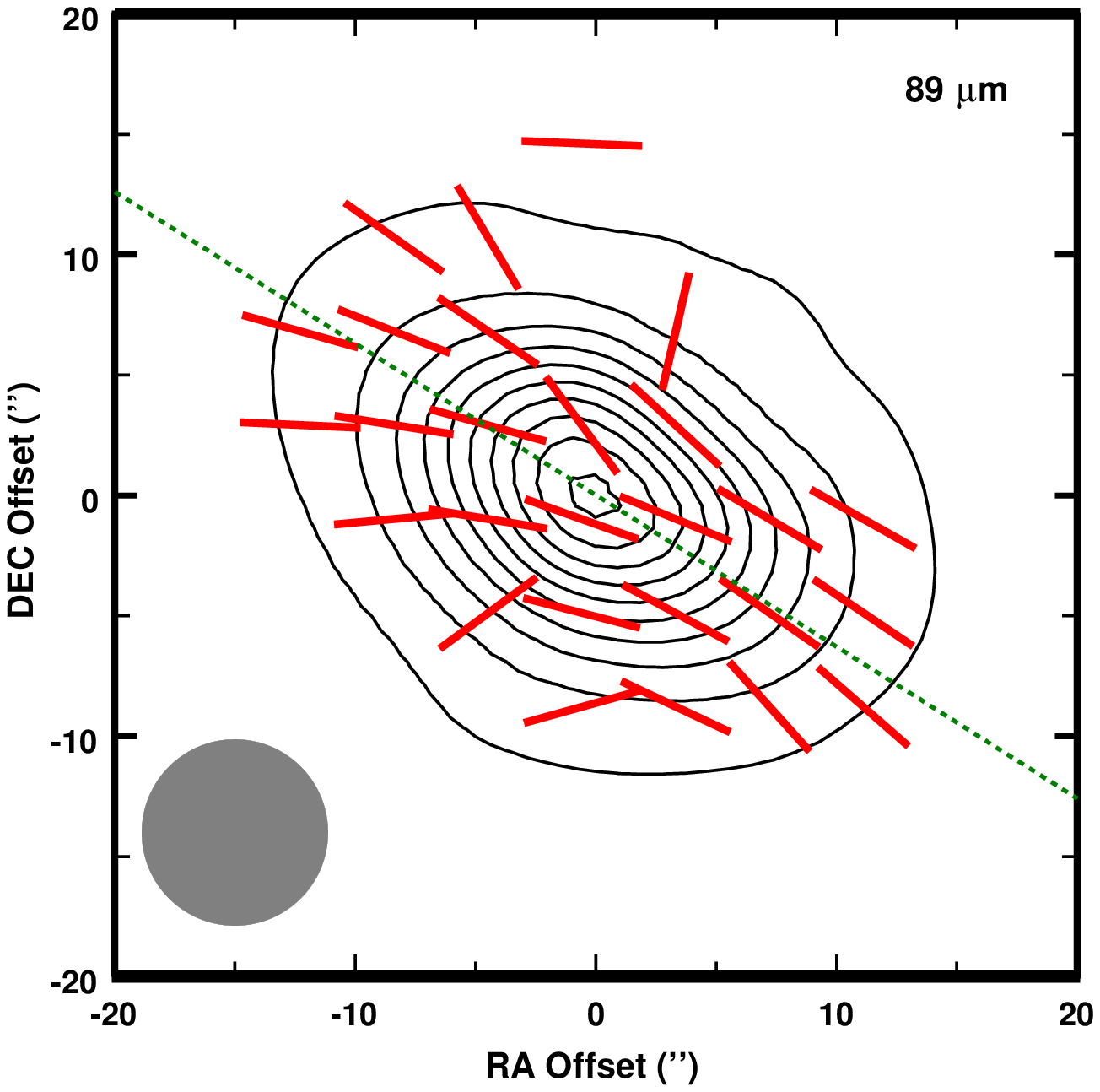}
\caption{Polarization vector map of NGC 253 at $89~\micron$, with vectors rotated $90\degr$ to represent the inferred magnetic field direction. Left: Vectors with length proportional to the fractional polarization. Right: Position angle only, overlaying intensity contours. The first contour starts at $2.8 \times 10^4$ MJy/sr with increments of $2.8 \times 10^4$ MJy/sr. The dashed line indicates the long axis of the tilted disk \citep[PA $= 51 \degr$,][]{penc80}.}
\end{center}
\end{figure}

The SOFIA/HAWC+ 89\micron\ observations of NGC 253 are shown in Figure 5, where we have plotted polarization vectors on a grid with half beam--width for the spacing and with position angles rotated $90\degr$ to represent the inferred magnetic field direction. M82 (d = 3.6Mpc) and NGC 253 (d = 3.5Mpc) are at very similar distances, so our maps in RA and DEC are on nearly the same physical scale.  On the left, the vector length is proportional to the fractional polarization. On the right, all vectors are plotted with the same length to better clarify the position angle morphology. Cuts in fractional polarization are at the same S/N (3.3/1 debiased) as for M82, but with a cut at intensity contour of 0.38 Jy/\sq\arcsec at a wavelength of $89~\micron$. The polarization fraction ranges from a high of 2\% well off the nucleus to 0.1 to 0.2\% on the nucleus. The polarization at the nucleus is below our nominal systematic error of 0.3\%, but the position angle is consistent with the vectors to the NE and SW along the major axis, not a vertical geometry. Unlike M82, the rotated polarization vectors lie largely along the long axis of the tilted disk in NGC 253. However, there is some evidence for a vertical field geometry above and below the plane to the NW and SE along the minor axis. 

Radio synchrotron observations of NGC 253 \citep{hees09} show a magnetic field geometry that consists of disk and halo components. The disk component dominates the visible disk with the magnetic field orientation parallel to the disk at small distances from the midplane. Well out in the halo, the field (as measrued at radio wavelengths) shows the familiar X--shape seen in several nearly edge--on galaxies \citep{beck13}. The radio observations show no indication of a vertical component along the minor axis of the tilted disk. We do find several vectors away from the plane (dotted line in Figure 5) that might be indicative of a verctical component, but such a vertical field is much more obvious in M82. Compared to M82, NGC 253 can not have as large a fraction of the dust column depth containing a vertical field.  

Figure 6 plots the trend of fractional polarization with surface brightness, similar to Figure 4 for M82, except we are using surface brightness as a proxy for optical depth. Upper limits are plotted as green triangles. The data show a decline in polarization with intensity, similar to M82. A rough fit to the slope of this trend is much steeper than $P\propto I^{-1/2}$. The steeper decline in polarization with intensity seen in NGC 253 must be due to greater large-scale cancellation effects with column depth in this galaxy. As with M82, if a `crossed--field' effect is at work, and it spans most of the area we have mapped, then it will lower the overall fractional polarization. Since none of the vectors near the nucleus along the disk show a vertical geometry, any `crossed--field' effect taking place must be dominated by the polarization in the disk, not the wind. For M82 we were able to use the NIR polarimetry to roughly estimate the fraction of the column depth that had vertical and planar fields. We do not have NIR polarimetry of NGC 253, but by analogy, the low net polarization in the disk of NGC 253 would indicate roughly 2/3 of the column depth contains a planar field, with no more than 1/3 associated with a vertical field and the super--galactic wind in this galaxy.

\begin{figure}
\begin{center}
\includegraphics[width=5.0in]{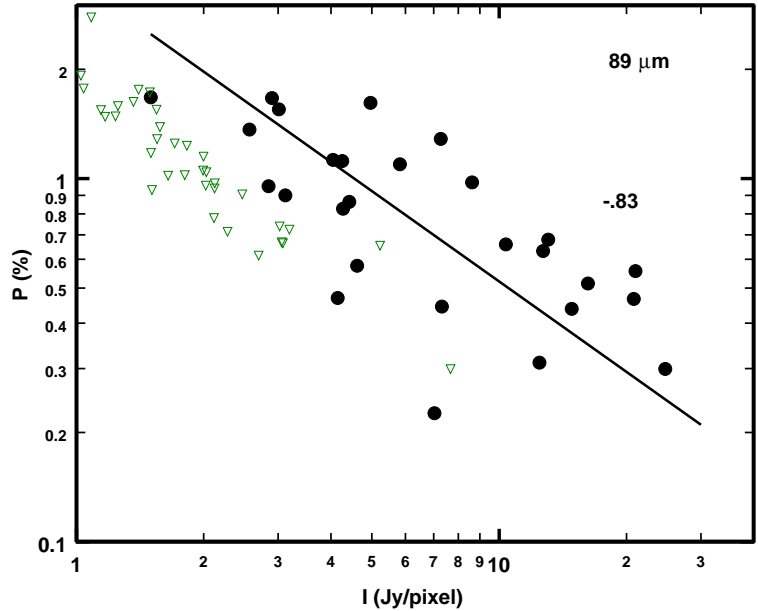}
\caption{Plot of fractional polarization against surface brightness for high S/N data points at $89~\micron$ for NGC 253. The intensity is being used as a proxy for optical depth. $3.3 \sigma$ Upper limits are shown as green trangles.}
\end{center}
\end{figure}

\section{Discussion, Integrated Properties} \label{sec:disc}

The underlying mechanism for producing massive winds from the central regions of starburst galaxies is not understood. Detection of the wind of M82 inspired the pioneering work of \cite{chevalier85} on thermally driven winds. Later works explored the role of radiation pressure \citep[e.g.][]{murr11,krumholz18}, cosmic ray driving \citep[e.g.][]{everett08}, and combinations of these effects \citep{hopkins12,ruszkowski17}. Thermal models show a tight correlation between central temperature and asymptotic velocity, at least when the wind is sufficiently hot and/or tenuous that radiative cooling is insignificant \citep{bustard17}. Thus, $\sim$10$^8$K gas is required to reach the speeds of up to 2200 km s$^{-1}$ detected in the x-ray emitting gas in the wind of M82 \citep{strickland07}. The origin of the cooler and slower gas observed in the outflow is unclear, although it may form \textit{in situ} through shock compression in the flow
or through the effects of repeated supernova explosions driving supershells in the central regions of molecular disks \citep{fuji09}. 

It is expected that a wind, whatever its origin, as massive as the one in M82 will drag the magnetic fieldlines out along with it. In this sense, the transition to a near vertically-oriented field in the starburst core of M82 is not surprising. However, given that both the spectroscopic and imaging evidence for the wind is in warm to hot ionized gas, it is notable that the field is vertical in the warm dust, which presumably is situated in mostly molecular gas associated with star forming regions (but not dense cores). This  suggests that the clouds and the intercloud medium are magnetically connected and that the field in the clouds is not overwhelmingly tangled by turbulence. \cite{walt02} find kinematic signatures of an outflow in observations of the molecular gas in M82, a characteristic in common with NGC 253 \citep{bolatto13}. It is interesting that the observations reported in this paper show weaker evidence for a vertical field in NGC 253, however.

Polarized emission from aligned dust grains provides information on the magnetic field geometry in the interstellar medium but does not directly measure magnetic field strength, and primarily traces the field in both the diffuse and molecular gas. Indirect methods of measuring field strength such as the Chandraskhar-Fermi method \citep{CF} or measures of the dispersion in position angle \citep[see][and references therin]{houd16}, cannot be applied here, as there  is simply too much averaging taking place in our 90pc beam. M82 is a strong emitter of both nonthermal radio radiation and $\gamma$-rays, and modeling their spectra can provide an estimate of the mean magnetic field strength \citep{yoas13}. Their best fit model  assumes a total molecular gas mass of $\sim 4\times 10^8M_{\odot}$, a factor of 10 larger than the mass estimated here, to be threaded by the vertical field, and yields $B \sim 250~\mu {\rm{G}}$ and a wind speed of 500 km s$^{-1}$. Their derived magnetic field strength is somewhat above the field strength of $150~\mu {\rm{G}}$ used by \cite{dece09} under the assumption that the magnetic field energy density is in equipartition with the cosmic ray energy density. A range of models with larger fields and faster winds or smaller fields and slower winds fits the data nearly as well. The field strength is weighted by ISM properties in a complex way, with most of the synchrotron radiation being emitted in the low--density, large--filling--factor medium, but most of $\gamma$-rays and secondary leptons are produced in the high density clouds. A similar modeling attempt for NGC 253 \citep{yoas14} failed to produce a good joint fit for both the radio and $\gamma$-ray spectra. Whether the different outcomes for M82 and NGC 253 are related in any way to the different polarization properties reported here is beyond the scope of this paper.

The Galactic Center of the Milky Way also has a vertical magnetic field geometry immediately above and below the inner disk (which itself has a planar field) as evidended by the presence of numerous magnetized vertical filaments \citep{yuse04, morr06}. If the magnetic field strength in these filaments is at the upper end of the range allowed by measurements, $\sim 1~$mG, stronger than model estimates for the wind in M82, then the vertical field dominates gas dynamics. Perhaps the vertical field in the Galactic Center could have an origin unrelated to winds, unlike our interpretation of the field geometry in M82. Note that if placed at the distance of M82, FIR polarimetry of the Galactic Center would likely be dominated by emission from dense molecular clouds with a planar field geometry \citep[e.g. Fig. 1 in][]{chus05}.

M82 and NGC 253 are nearby galaxies that allow us to map polarized dust emission on $80-100$pc scales. For comparison with future observations of more distant galaxies, we have computed the integrated properties of these two galaxies in much larger beams. The results are shown in Table~\ref{tbl:integrated}. Note that the net position angles from the integrated I, Q and U maps for M82 at $53~\micron$ and NGC 253 at $89~\micron$ preserve information on the magnetic field geometry relative to the disk and wind position angles. This implies that future observations of at least some more distant, unresolved (at FIR wavelengths) galaxies with known jet, wind, or disk geometries can still provide relevant information on the global magnetic field geometry.  However, the large--scale planar and vertical components in the $154~\micron$ map of M82 cancel and produce very low fractional polarization in the integrated signal.

\begin{deluxetable}{ccccc}
\caption{Integrated Properties}
\label{tbl:integrated}
\tablewidth{0pt}
\tablehead{\colhead{Map} & \colhead{radius $(\arcsec)$} & \colhead{I (Jy)} & \colhead{P (\%)} & \colhead{PA $\left( \vec{E}{\rm{,}}~\degr \right)$ }} 
\startdata
M82 $53~\micron$ & 45 & 2110 & 1.5 & 81 \\
M82 $154~\micron$ & 60 & 1430 & 0.1 & -- \\
NGC253 $53~\micron$ & 15 & 960 & 0.1 & -- \\
NGC253 $89~\micron$ & 30 & 1450 & 0.4 & 160 \\
\enddata
\end{deluxetable}

\section{Conclusions}

We have presented FIR polarimetric imaging observations of M82 and NGC 253 using HAWC+ on SOFIA.  Effects such as scattering seen at NIR wavelengths and Faraday rotation at radio frequencies are absent in the FIR polarimetry. Unlike radio synchrotron emission, the FIR emission is sensitive to the dust column density (weighted by temperature) along the line of sight, not the population of relativistic electrons. These observations of M82 are consistent with a vertical magnetic field in the central $40\arcsec \times 20\arcsec$ region where the starburst is located and a planar magnetic field in the surrounding disk. The fractional polarization is very low at the nucleus, but shows the same vertical field geometry. The low polarization could be due to the mutually perpendicular nuclear and disk fields partially canceling the polarization or this effect in combination with strong turbulence on scales much smaller than our beam. If the `crossed--field' effect dominates, then $\sim 5-6 \times 10^7~ \rm{M_\odot}$ in the central region of M82 is threaded with a magnetic field perpendicular to the disk. For NGC 253 the observations at $89~\micron$ are consistent with a planar magnetic field geometry both in nucleus and to the NE and SW along the disk. There is some indication of a more vertical geometry along the minor axis, but off the nucleus. Compared to M82, NGC 253 can not have as much of the dust column depth containing a vertical field.

\subsection{Acknowledgments}

\acknowledgments

Based [in part] on observations made with the NASA/DLR Stratospheric Observatory for Infrared Astronomy (SOFIA). SOFIA is jointly operated by the Universities Space Research Association, Inc. (USRA), under NASA contract NAS2-97001, and the Deutsches SOFIA Institut (DSI) under DLR contract 50 OK 0901 to the University of Stuttgart.  Part of this research was carried out at the Jet Propulsion Laboratory, California Institute of Technology, under a contract with the National Aeronautics and Space Administration. The LIC code was ported from publicly-available IDL source by Diego Falceta-Gon\c{c}alves.

Herschel is an ESA space observatory with science instruments provided by European-led Principal Investigator consortia and with important participation from NASA.

\vspace{5mm}
\facilities{SOFIA}


\begin{thebibliography}{}

\bibitem[Andersson et al.(2015)]{ande15} Andersson, B.-G., Lazarian, A., \& Vaillancourt, J.~E.\ 2015, \araa, 53, 501 
\bibitem[Adebahr et al.(2017)]{adeb17} Adebahr, B., Krause, M., Klein, U., Heald, G., \& Dettmar, R.-J.\ 2017, \aap, 608, A29 
\bibitem[Beck(2015)]{beck15} Beck, R.\ 2015, \aapr, 24, 4 
\bibitem[Beck \& Wielebinski(2013)]{beck13} Beck, R., \& Wielebinski, R.\ 2013, Planets, Stars and Stellar Systems.~Volume 5: Galactic Structure and Stellar Populations, 5, 641 
\bibitem[Bertone et al.(2006)]{bert06} Bertone, S., Vogt, C., \& En{\ss}lin, T.\ 2006, \mnras, 370, 319 
\bibitem[Bolatto et al.(2013)]{bolatto13} Bolatto, A.~D., Warren, S.~R., Leroy, A.~K., et al.\ 2013, \nat, 499, 450 
\bibitem[Boselli et al.(2012)]{bose12} Boselli, A., Ciesla, L., Cortese, L., et al.\ 2012, \aap, 540, A54 
\bibitem[Brevik(2012)]{brevik12} Brevik, J.~A.\ 2012, Ph.D.~Thesis, California Institute of Technology
\bibitem[Bustard et al.(2017)]{bustard17} Bustard, C., Zweibel, E.~G., \& Cotter, C.\ 2017, \apj, 835, 72 
\bibitem[Cabral \& Leedom(1993)]{cabr93} Cabral, B., \& Leedom, L.C.\ 1993, in Proceedings of the 20th annual conference on Computer graphics and interactive techniques, ACM, 263-270
\bibitem[Chandrasekhar \& Fermi(1953)]{CF} Chandrasekhar, S., \& Fermi, E.\ 1953, \apj, 118, 113 
\bibitem[Chevalier \& Clegg(1985)]{chevalier85} Chevalier, R.~A., \& Clegg, A.~W.\ 1985, \nat, 317, 44 
\bibitem[Chuss et al.(2018)]{chus18} Chuss, D.~T., Andersson, B-G, Bally, J., et al.\ 2018, \apj, submitted
\bibitem[Chuss et al.(2005)]{chus05} Chuss, D.~T., Dowell, C.~D., Hildebrand, R.~H., \& Novak, G.\ 2005, Astronomical Polarimetry: Current Status and Future Directions, 343, 311 
\bibitem[Davidge(2008)]{davi08} Davidge, T.~J.\ 2008, \aj, 136, 2502 
\bibitem[de Cea del Pozo et al.(2009)]{dece09} de Cea del Pozo, E., Torres, D.~F., \& Rodriguez Marrero, A.~Y.\ 2009, \apj, 698, 1054 
\bibitem[de Vaucouleurs(1958)]{deva58} de Vaucouleurs, G.\ 1958, \apj, 127, 487 
\bibitem[Dowell et al.(2010)]{dowe10} Dowell, C.~D., Cook, B.~T., Harper, D.~A., et al.\ 2010, \procspie, 7735, 77356H 
\bibitem[Draine \& Lee(1984)]{drai84} Draine, B.~T., \& Lee, H.~M.\ 1984, \apj, 285, 89 
\bibitem[Engelbracht et al.(2006)]{enge06} Engelbracht, C.~W., Kundurthy, P., Gordon, K.~D., et al.\ 2006, \apjl, 642, L127 
\bibitem[Everett et al.(2008)]{everett08} Everett, J.~E., Zweibel, E.~G., Benjamin, R.~A., et al.\ 2008, \apj, 674, 258 
\bibitem[F{\"o}rster Schreiber et al.(2003)]{fors03} F{\"o}rster Schreiber, N.~M., Genzel, R., Lutz, D., \& Sternberg, A.\ 2003, \apj, 599, 193 
\bibitem[Fujita et al.(2009)]{fuji09} Fujita, A., Martin, C.~L., Mac Low, M.-M., New, K.~C.~B., \& Weaver, R.\ 2009, \apj, 698, 693 
\bibitem[Greaves et al.(2000)]{grea00} Greaves, J.~S., Holland, W.~S., Jenness, T., \& Hawarden, T.~G.\ 2000, \nat, 404, 732
\bibitem[Greco et al.(2012)]{grec12} Greco, J.~P., Martini, P., \& Thompson, T.~A.\ 2012, \apj, 757, 24 
\bibitem[Griffin et al.(2010)]{grif2010} Griffin, M.~J., Abergel, A., Abreu, A., et al.\ 2010, \aap, 518, L3 
\bibitem[Harper et al.(2018)]{harp18} Harper, D.~A., Runyan, M.~C., Dowell, C.~D., et al.\ 2018, Journal of Astronomical Instrumentation, accepted
\bibitem[Harper \& Low(1973)]{harp73} Harper, D.~A., Jr., \& Low, F.~J.\ 1973, \apjl, 182, L89 
\bibitem[Heesen et al.(2009)]{hees09} Heesen, V., Krause, M., Beck, R., \& Dettmar, R.-J.\ 2009, \aap, 506, 1123 
\bibitem[Helou et al.(1985)]{helo85} Helou, G., Soifer, B.~T., \& Rowan-Robinson, M.\ 1985, \apjl, 298, L7 
\bibitem[Hildebrand(1983)]{hild83} Hildebrand, R.~H.\ 1983, \qjras, 24, 267 
\bibitem[Hildebrand et al.(2009)]{hild09} Hildebrand, R.~H., Kirby, L., Dotson, J.~L., Houde, M., \& Vaillancourt, J.~E.\ 2009, \apj, 696, 567 
\bibitem[Hildebrand et al.(1999)]{hild99} Hildebrand, R.~H., Dotson, J.~L., Dowell, C.~D., Schleuning, D.~A., \& Vaillancourt, J.~E.\ 1999, \apj, 516, 834 
\bibitem[Hopkins et al.(2012)]{hopkins12} Hopkins, P.~F., Quataert, E., \& Murray, N.\ 2012, \mnras, 421, 3522 
\bibitem[Houde et al.(2016)]{houd16} Houde, M., Hull, C.~L.~H., Plambeck, R.~L., Vaillancourt, J.~E., \& Hildebrand, R.~H.\ 2016, \apj, 820, 38 
\bibitem[Jones(1989)]{jone89} Jones, T.~J.\ 1989, \apj, 346, 728 
\bibitem[Jones(1993)]{jone93} Jones, T.~J.\ 1993, \apj, 403, 135 
\bibitem[Jones(2000)]{jone00} Jones, T.~J.\ 2000, \aj, 120, 2920 
\bibitem[Jones et al.(2015)]{joba15} Jones, T.~J., Bagley, M., Krejny, M., Andersson, B.-G., \& Bastien, P.\ 2015, \aj, 149, 31 
\bibitem[Jones \& Whittet(2015)]{jowh15} Jones, T.~J., \& Whittet, D., C.~B.\ 2015, Polarimetry of Stars and Planetary Systems, 147 
\bibitem[Kaneda et al.(2010)]{kane10} Kaneda, H., Ishihara, D., Suzuki, T., et al.\ 2010, \aap, 514, A14 
\bibitem[Karachentsev \& Kashibadze(2006)]{kara06} Karachentsev, I.~D., \& Kashibadze, O.~G.\ 2006, Astrophysics, 49, 3 
\bibitem[Kobulnicky et al.(1994)]{kobu94} Kobulnicky, H.~A., Molnar, L.~A., \& Jones, T.~J.\ 1994, \aj, 107, 1433 
\bibitem[Kronberg et al.(1999)]{kron99} Kronberg, P.~P., Lesch, H., \& Hopp, U.\ 1999, \apj, 511, 56 
\bibitem[Krumholz et al.(2018)]{krumholz18} Krumholz, M.~R., Burkhart, B., Forbes, J.~C., \& Crocker, R.~M.\ 2018, \mnras, 477, 2716 
\bibitem[Larkin et al.(1994)]{lark94} Larkin, J.~E., Graham, J.~R., Matthews, K., et al.\ 1994, \apj, 420, 159 
\bibitem[Lim et al.(2013)]{lim13} Lim, S., Hwang, N., \& Lee, M.~G.\ 2013, \apj, 766, 20 
\bibitem[Martini et al.(2018)]{mart18} Martini, P., Leroy, A.~K., Mangum, J.~G., et al.\ 2018, \apj, 856, 61 
\bibitem[Matthews et al.(2009)]{matt09} Matthews, B.~C., McPhee, C.~A., Fissel, L.~M., \& Curran, R.~L.\ 2009, \apjs, 182, 143
\bibitem[McLeod et al.(1993)]{mcle93} McLeod, K.~K., Rieke, G.~H., Rieke, M.~J., \& Kelly, D.~M.\ 1993, \apj, 412, 111 
\bibitem[Morris(2006)]{morr06} Morris, M.\ 2006, Journal of Physics Conference Series, 54, 1 
\bibitem[Murray et al.(2011)]{murr11} Murray, N., M{\'e}nard, B., \& Thompson, T.~A.\ 2011, \apj, 735, 66 
\bibitem[Naylor et al.(2010)]{nayl10} Naylor, B.~J., Bradford, C.~M., Aguirre, J.~E., et al.\ 2010, \apj, 722, 668 
\bibitem[Nikola et al.(2012)]{niko12} Nikola, T., Herter, T.~L., Vacca, W.~D., et al.\ 2012, \apjl, 749, L19 
\bibitem[Ohyama et al.(2002)]{ohym02} Ohyama, Y., Taniguchi, Y., Iye, M., et al.\ 2002, \pasj, 54, 891 
\bibitem[Pence(1980)]{penc80} Pence, W.~D.\ 1980, \apj, 239, 54 
\bibitem[P{\'e}rez-Beaupuits et al.(2018)]{pere18} P{\'e}rez-Beaupuits, J.~P., G{\"u}sten, R., Harris, A., et al.\ 2018, \apj, 860, 23 
\bibitem[Pilbratt et al.(2010)]{pilb2010} Pilbratt, G.~L., Riedinger, J.~R., Passvogel, T., et al.\ 2010, \aap, 518, L1 
\bibitem[Planck Collaboration et al.(2018)]{plan18} Planck Collaboration, Aghanim, N., Akrami, Y., et al.\ 2018, arXiv:1807.06212 
\bibitem[Rekola et al.(2005)]{reko05} Rekola, R., Richer, M.~G., McCall, M.~L., et al.\ 2005, \mnras, 361, 330 
\bibitem[Reuter et al.(1994)]{reut94} Reuter, H.-P., Klein, U., Lesch, H., Wielebinski, R., \& Kronberg, P.~P.\ 1994, \aap, 282, 724 
\bibitem[Rieke et al.(1980)]{riek80} Rieke, G.~H., Lebofsky, M.~J., Thompson, R.~I., Low, F.~J., \& Tokunaga, A.~T.\ 1980, \apj, 238, 24 
\bibitem[Roussel et al.(2010)]{rous10} Roussel, H., Wilson, C.~D., Vigroux, L., et al.\ 2010, \aap, 518, L66 
\bibitem[Ruszkowski et al.(2017)]{ruszkowski17} Ruszkowski, M., Yang, H.-Y.~K., \& Zweibel, E.\ 2017, \apj, 834, 208 
\bibitem[Sharp \& Bland-Hawthorn(2010)]{shar10} Sharp, R.~G., \& Bland-Hawthorn, J.\ 2010, \apj, 711, 818 
\bibitem[Shopbell \& Bland-Hawthorn(1998)]{shop98} Shopbell, P.~L., \& Bland-Hawthorn, J.\ 1998, \apj, 493, 129 
\bibitem[Spitzer(1978)]{spit78} Spitzer, L.\ 1978, Physical processes in the interstellar medium, by Lyman Spitzer.~ New York Wiley-Interscience, 1978.~333 p.,  
\bibitem[Strickland \& Heckman(2007)]{strickland07} Strickland, D.~K., \& Heckman, T.~M.\ 2007, \apj, 658, 258 
\bibitem[Telesco(1988)]{tele88} Telesco, C.~M.\ 1988, \araa, 26, 343 
\bibitem[Telesco et al.(1989)]{tele89} Telesco, C.~M., Decher, R., \& Joy, M.\ 1989, \apjl, 343, L13 
\bibitem[Uhlig et al.(2012)]{uhli12} Uhlig, M., Pfrommer, C., Sharma, M., et al.\ 2012, \mnras, 423, 2374 
\bibitem[Vaillancourt et al.(2007)]{vail07} Vaillancourt, J.~E., Chuss, D.~T., Crutcher, R.~M., et al.\ 2007, \procspie, 6678, 66780D 
\bibitem[Veilleux et al.(2005)]{veil05} Veilleux, S., Cecil, G., \& Bland-Hawthorn, J.\ 2005, \araa, 43, 769 
\bibitem[Voelk(1989)]{voel89} Voelk, H.~J.\ 1989, \aap, 218, 67 
\bibitem[Wardle \& Kronberg(1974)]{ward74} Wardle, J.~F.~C., \& Kronberg, P.~P.\ 1974, \apj, 194, 249 
\bibitem[Walter et al.(2017)]{walt17} Walter, F., Bolatto, A.~D., Leroy, A.~K., et al.\ 2017, \apj, 835, 265 
\bibitem[Walter et al.(2002)]{walt02} Walter, F., Weiss, A., \& Scoville, N.\ 2002, \apjl, 580, L21 
\bibitem[Yoast-Hull et al.(2014)]{yoas14} Yoast-Hull, T.~M., Gallagher, J.~S., III, Zweibel, E.~G., \& Everett, J.~E.\ 2014, \apj, 780, 137 
\bibitem[Yoast-Hull et al.(2013)]{yoas13} Yoast-Hull, T.~M., Everett, J.~E., Gallagher, J.~S., III, \& Zweibel, E.~G.\ 2013, \apj, 771, 73 
\bibitem[Yusef-Zadeh et al.(2004)]{yuse04} Yusef-Zadeh, F., Hewitt, J.~W., \& Cotton, W.\ 2004, \apjs, 155, 421 
\bibitem[Zweibel(1996)]{zwei96} Zweibel, E.~G.\ 1996, Polarimetry of the Interstellar Medium, 97, 486 


\end{thebibliography}
\end{document}